\documentstyle[11pt]{article}
\newcommand{\blankline}{\vskip .3cm}
\newcommand{\f}{\begin{equation}}
\newcommand{\ff}{\end{equation}}

\begin{document}
\centerline{\LARGE  A possible solution to the problem of time in
quantum cosmology}
\blankline
\rm
\centerline{Stuart Kauffman${}^\dagger$ and Lee Smolin${}^*$}
\blankline
\centerline{\it ${}^\dagger$  Santa Fe Institute}
\centerline{\it 1399 Hyde Park Road, Santa Fe, New Mexico, 87501-8943}
\centerline{\it  ${}^*$
Center for Gravitational Physics and Geometry}
\centerline{\it Department of Physics}
 \centerline {\it The Pennsylvania State University}
\centerline{\it University Park, PA, USA 16802}
 \vfill
\centerline{March 5, 1997}
\vfill
\centerline{ABSTRACT}
 
We argue that in classical and quantum theories of
gravity the configuration space and Hilbert space may
not be constructible through any finite procedure.  If this is the
case then the ``problem of time" in quantum cosmology
may be a pseudo-problem, because the argument that time
disappears from the theory depends on constructions that cannot be realized
by any finite beings that live in the universe.  We propose an alternative
formulation of quantum cosmological theories in which it is only
necessary to predict the amplitudes for any given state to 
evolve to a finite number of possible successor states.  The space
of accessible states of the system is then constructed as the
universe evolves from any initial state.  In this kind of formulation
of quantum cosmology time and causality are built in at the 
fundamental level.  An example of such a theory is the recent path
integral formulation of quantum gravity of Markopoulou and Smolin, 
but there are a wide class of theories of this type.

\blankline
${}^*$ smolin@phys.psu.edu
\eject

\section{Introduction}

The problem of time in quantum cosmology is one of the key conceptual
problems faced by theoretical physics at the present time.  Although
it was first raised during the 1950's, it has resisted solution, 
despite many different kinds of 
attempts\cite{chris-review,karel-review,Carlo-time,spaceandtime,Julian-time}.  
Here we would like
to propose a new kind of approach to the problem.  Basically, we will
argue that the problem is not with time, but with some of the assumptions
that lead to the conclusion that there is a problem.  These are assumptions
that are quite satisfactory in ordinary quantum mechanics, but that
are problematic in quantum gravity, because they may not be realizable
with any constructive procedure.  In a quantum theory of cosmology this
is a serious problem, because one wants any theoretical construction
that we use to describe the universe to be something that can be
realized in a finite time, by beings like ourselves that live in
that universe.  If the quantum theory of cosmology requires a
non-constructible procedure to define its formal setting, it is
something that could only be of use to a mythical creature of infinite
capability.  One of the things we would like to demand of a quantum
theory of cosmology is that it not make any reference to anything at
all that might be posited or imagined to exist outside the closed
system which is the universe itself.

We believe that this requirement has a number of consequences for
the problem of constructing quantum a good quantum theory of
cosmology.  These have been discussed in detail elsewhere
\cite{spaceandtime,book,investigations}.  
Here we would like to describe one more implication of
the requirement, which appears to bear on the problem of time.

We begin by summarizing briefly the argument that time is not
present in a quantum theory of cosmology.  In section 3 we 
introduce a worry 
that one of the assumptions of the argument may not be realizable
by any finite procedure.  (Whether this is actually the case is 
not known presently.)  We explain how the argument for the disappearance
of time would be affected by this circumstance.  Then we explain how
a quantum theory of cosmology might be made which overcomes the
problem, but at the cost of introducing a notion of time and
causality at a fundamental level.  As an example we refer to recent
work on the path integral for quantum gravity\cite{causal}, but the
form of the theory we propose is more general, and may apply
to a wide class of theories beyond quantum general relativity.

\section{The argument for the absence of time}

The argument that time is not a fundamental aspect of the
world goes like this\footnote{For more details and
discussion see 
\cite{chris-review,karel-review,Carlo-time,spaceandtime,Julian-time}}.  
In classical mechanics one begins with
a space of configurations $\cal C$
of a system $\cal S$.  Usually the system $\cal S$ is assumed to
be a subsystem of the universe.  In this case there is a clock
outside the system,  which is carried by some inertial observer.
This clock is used to label the trajectory of the system in the
configuration space $\cal C$.  The classical trajectories are
then extrema of some action principle, $\delta I =0$.  

Were it not for the external clock, one could already say that
time has disappeared, as each trajectory exists all at once
as a curve $\gamma$ on $\cal C$.  Once the trajectory is
chosen, the whole history of the system is determined.
In this sense there is nothing in the description that corresponds
to what we are used to thinking of as a flow or progression of
time.  Indeed, just as the whole of any one trajectory exists
when any point and velocity are specified, the whole set
of trajectories may be said to exist as well, as a timeless
set of possibilities.

Time is in fact represented in the description, but it is not in any
sense a time that is associated with the system itself.  Instead, the
$t$ in ordinary classical mechanics refers to a clock carried by
an inertial observer, which is not part of the dynamical system
being modeled.  This 
external clock is represented in the configuration space
description as a special parameterization of each
trajectory, according to which the equations of motion
are satisfied.  Thus, it may be said that there is no sense in which
time as something physical is represented in classical mechanics,
instead the problem is postponed, as what is represented is time
as marked by a clock that exists outside of the physical system
which is modeled by the trajectories in the configuration
space $\cal C$.

In quantum mechanics the situation is rather similar.  There
is a $t$ in the quantum state and the Schroedinger equation,
but it is time as measured by an external clock, which is
not part of the system being modeled.  
Thus, when we write,
\f
\imath \hbar {d \over dt} \Psi (t) = \hat{H} \Psi (t)
\label{sch}
\ff
the Hamiltonian refers to evolution, as it would be 
measured by an external observer, who refers to the external
clock whose reading is $t$

The quantum state
can be represented as a function $\Psi$ over the configuration
space, which is normalizable in some inner product.  The inner
product is another a priori structure, it refers also to the external
clock, as it is the structure that allows us to represent the
conservation of probability as measured by that clock.

When we turn to the problem of constructing a cosmological
theory we face a key problem, which is that there is no
external clock.  There is by definition nothing outside of the system,
which means that the interpretation of the theory must be
made without reference to anything that is not part of the
system which is modeled.
In classical cosmological theories, such as general relativity applied
to spatially compact universes, or models such as the Bianchi
cosmologies or the Barbour-Bertotti model\cite{BB1,BB-Royal}, 
this is expressed by
the dynamics having a gauge invariance, which includes 
arbitrary reparameterizations of the classical trajectories.
(In general relativity this is part of the diffeomorphism invariance
of the theory.)  As a result, the classical theory is expressed in
a way that makes no reference to any particular 
parameterization of the
trajectories. Any parameterization is as good as any other,
none has any physical meaning.   
The solutions are then labeled by a trajectory,
$\gamma$, period, there is no reference to a parameterization.

This is the sense in which time may be said to disappear from
classical cosmological theories.  There is nothing in the theory that
refers to any time at all.  At least without a good deal more
work, the theory speaks only in terms of the whole
history or trajectory, it seems to have nothing to say about what the
world is like at a particular moment.  

There is one apparently straightforward way out of this, which is to
try to define an intrinsic notion of time, in terms of physical 
observables.  One may construct parameter independent observables
that describe what is happening at a point on the trajectory if that
point can be labeled intrinsically by some physical property.  For
example, one might consider some particular 
degree of freedom to be an intrinsic,
physical clock, and label the points on the trajectory by its value.
This works in some model systems, but in interesting cases such
as general relativity it is not known if such an intrinsic notion of
time exists which is well defined over the whole of the configuration
space.

In the quantum theory there is a corresponding phenomenon.  As there
is no external $t$ with which to measure evolution of the quantum
state one has instead of (\ref{sch}) the quantum constraint
equation
\f
\hat{H} \Psi =0
\label{constraint}
\ff
where $\Psi$ is now just a function on the configuration space.
Rather than describing evolution, eq. (\ref{constraint}) generates 
arbitrary parameterizations of the trajectories.
The wavefunction must be normalizable under an inner product,
given by some density $\rho$ on the configuration space.  The
space of physical states is then given by (\ref{constraint}) subject
to
\f
1 = \int_{\cal C}\rho \bar{\Psi}\Psi
\label{ip}
\ff

We see that, at least naively
time has completely disappeared from the formalism.  This has led to
what is called the ``problem of time in quantum cosmology", which
is how to either A) find an interpretation of the theory that restores
a role for time or B) provide an interpretation according to which
time is not part of a fundamental description of the world, but only
reappears in an appropriate classical limit.

There have been various attempts at either direction.  We will not
describe them here, except to say that, in our opinion, 
so far none has proved
completely satisfactory\footnote{For good critical reviews
that deflate most known proposals, see 
\cite{chris-review,karel-review}.}.  There are a number of attempts at A)
which succeed when applied to either models or the semiclassical
limit, but it is not clear whether any of them overcome 
technical obstacles of various kinds when applied to the full theory.
The most well formulated attempt of type B), which is 
that of Barbour\cite{Julian-time},
may very well be logically consistent.   But it forces one to swallow
quite a radical point of view about the relationship between time
and our experience.

Given this situation, we would like to propose that the problem may
be not with time, but with the assumptions of the argument that
leads to time being absent.  Given the number of attempts that
have been made to resolve the problem, 
which have not so far led to a good solution, perhaps it might
be better to try to dissolve the issue by questioning one of
the assumptions of the argument that leads to the statement of
the problem.
This is what we would like to do in the following.

\section{A problem with the argument for the 
disappearance of time}

Both the classical and quantum mechanical versions of the
argument for the disappearance of time begin with the
specification of the classical configuration space $\cal C$.
This seems an innocent enough assumption.  For a system
of $N$ particles in $d$ dimensional Euclidean space, it is simply 
$R^{Nd}$.  One can then find the corresponding basis of
the Hilbert space by simply enumerating the Fourier modes.
Thus, for cases such as this, it is certainly the case that the
configuration space and the Hilbert space structure can be
specified {\it a priori.}

However, there are good reasons to suspect that
for cosmological theories it may not be so easy to specify
the whole of the configuration or Hilbert space.   For example,
it is known that the configuration spaces of theories that
implement relational notions of space are quite complicated.
One example is the Barbour-Bertotti model\cite{BB1,BB-Royal}, 
whose configuration
space consists of the relative distances between $N$ particles
in $d$ dimensional Euclidean space.  While it is presumably
specifiable in closed form, this configuration space is rather 
complicated, as it is the quotient of $R^{Nd}$ by the
Euclidean group in $d$ dimensions\cite{Julian-time}.  

The configuration space of compact three geometries is 
even more complicated, as it is the quotient of the space of
metrics by the diffeomorphism group.  It is known not to
be a manifold everywhere.  Furthermore, it has a preferred
end, where the volume of the universe vanishes.

These examples serve to show that
the configuration spaces of cosmological theories are not simple
spaces like $R^{Nd}$, but may be considerably complicated.
This raises a question: could there be a theory so complicated
that its space of configurations is not constructible
through any finite procedure?  For example, is it possible that the 
topology of
an infinite dimensional configuration space were not
finitely specifiable?  And were this the case, what would
be the implications for how we understand dynamics\footnote{There 
is an analogous issue in theoretical biology.  
The problem is that 
it does not appear that a
pre-specifiable   set of ``functionalities" exists
in biology, where pre-specifiable means a
compact description of an effective procedure to characterize ahead of
time, each member of the  set\cite{stubooks,investigations}.  This
problem seems to limit the possibilities of a formal framework for
biology in which there is a pre-specified space of states which describe
the functionalities of elements of a biological system.
Similarly, one may question whether it is in principle
possible in economic theory to give in advance an {\it a priori}
list of all the possible kinds of jobs, or goods or 
services\cite{stubooks}.}?

We do not know whether in fact the configuration space of
general relativity is finitely specifiable.  The problem is
hard because the physical configuration space is not the space
of three metrics.  It is instead the space of equivalence classes
of three metrics (or connections, in some formalisms) under
diffeomorphisms. The problem is that it is not 
known if there is any effective procedure which
will label the equivalence classes.  

One can in fact see this issue
in one approach to describing the configuration space, due to
Newman and Rovelli\cite{CarloTed}.  There the physical 
configuration space 
consists of the diffeomorphism equivalence classes of a set of three
flows on a three
manifold. (These come from the intersections of the
level surfaces of three functions.)  These classes
are partially characterized by the topologies of the flow lines
of the vector fields.   We may note that these flow lines may
knot and link, thus a part of the problem of specifying the
configuration space involves classifying the knotting and linking
among the flow lines.  

Thus, the configuration space of general relativity 
cannot be completely described
unless the possible ways that flow lines may knot and link in
three dimensions are finitely specifiable. It may be noted
that there is a decision procedure, due to Hacken, for
knots, although it is very cumbersome\cite{knotclass}.
However, it is not obvious that this is sufficient to
give a decision procedure for configurations in general
relativity, because there we are concerned with smooth
data.  In the smooth category the flow lines may knot and link
an infinite number of times in any bounded region.
The resulting knots may not be classifiable.  All that is known
is that knots with a finite number of crossings 
are classifiable.  If these is no decision procedure to classify
the knotting and linking of smooth flow lines then the points of
the configuration space of general relativity may not be distinguished
by any decision procedure.  This means that the configuration space is
not constructible by any finite procedure.  

When we turn from the classical to the quantum theory
the same issue arises.  First of all, if the configuration
space is not constructible through any finite procedure, then there
is no finite procedure to define normalizable wave functions on that space.
One might still wonder whether there is some constructible basis
for the theory.  Given the progress of the last few years
in quantum gravity we can investigate this question directly,
as we know more about
the space of quantum states of general relativity than we
do about the configuration space of the theory.  This is because it
has been shown that the space of spatially diffeomorphism
invariant states of the quantum gravitational field has a
basis which is in one to one correspondence with the
diffeomorphism classes of a certain set of embedded, labeled
graphs $\Gamma$, in a given three manifold $\Sigma$ 
\cite{volume1,sn1}.  These
are arbitrary graphs, whose edges are labeled by spins
and whose vertices are labeled by the distinct ways to combine
the spins in the edges that meet there quantum mechanically.
These graphs are called {\it spin networks}, they were 
invented originally by Roger Penrose\cite{sn-roger}, and then discovered
to play this role in quantum gravity\footnote{For a
review of these developments see \cite{future}.  These results
have also more recently been formulated as theorems in a
rigorous formulation of diffeomorphism invariant quantum
field theories\cite{rigorous,baez-sn}.}.

Thus, we cannot label all the basis elements of quantum
general relativity unless the diffeomorphism classes of the
embeddings of spin networks in a three manifold
$\Sigma$ may be classified.  But it is not known whether this
is the case.   The same procedure that classifies the knots
is not, at least as far as is known, extendible to the case of
embeddings of graphs.

What if it is the case that the diffeomorphism classes of the
embeddings of spin networks cannot be classified?  
While it may be possible to
give a finite procedure that generates all the embeddings of
spin networks, if they are not classifiable there will be no
finite procedure to tell if a given one produced is or is not
the same as a previous network in the list.
In this case there will be no finite procedure to 
write the completeness relation or expand a
given state in terms of the basis.  There will consequently
be no finite procedure to test whether an operator is
unitary or not.  Without being able to do any of these things,
we cannot really say that we have a conventional quantum
mechanical description.   If spin networks are not classifiable, then
we cannot construct the Hilbert space of quantum general
relativity.

In this case then the whole set up of the problem of time
fails.  If the Hilbert space of spatially diffeomorphism
invariant states is not constructible, 
then we cannot formulate a quantum theory of cosmology
in these terms.  
There may be something that corresponds to a 
``wavefunction of the universe" but it cannot be a vector
in a constructible Hilbert space.
Similarly, if the configuration space $\cal C$ of the
theory is not constructible, then we cannot describe the
quantum state of the universe in terms of a 
normalizable function on
$\cal C$.  

We may note that a similar argument arises for the path
integral formulations of quantum gravity.  
It is definitely known that four manifolds
are not classifiable; this means that  path integral
formulations of quantum gravity that include sums over topologies
are not constructible through a finite procedure\cite{hartle-class}.

Someone may object that these arguments have to do with quantum
general relativity, which is in any case unlikely to 
exist.
One might even like to use this problem as an argument
against quantum general relativity.
However, the argument only uses the kinematics of the theory,
which is that the configuration space includes diffeomorphism
and gauge invariant classes of some metric or 
connection.  It uses nothing
about the actual dynamics of the theory, nor does it 
assume anything about which matter fields are included.
Thus, the argument applies to a large 
class of theories, including supergravity.

\section{Can we do physics without a constructible
state space?}

What if it is the case that the Hilbert space of quantum gravity is
not constructible because embedded graphs in three space are not
classifiable?  How do we do physics?  We would like to argue
now that there is a straightforward answer to this question.
But it is one that necessarily involves the introduction of
notions of time and causality.

One model for how to do physics in the absence of a constructible
Hilbert space is seen in a recent 
formulation of the path integral for quantum
gravity in terms of spin networks by 
Markopoulou and Smolin\cite{causal}\footnote{This followed the
development of a Euclidean path integral by 
Reisenberger\cite{mike-paths}
and by Reisenberger and Rovelli\cite{RR}.  Very interesting
related work has also been done by John Baez\cite{jb-path}.
We may note that the theory described in \cite{causal} involves
non-embedded spin networks, which probably are classifiable, but
it can be extended to give a theory of the evolution of
embedded spin networks.}.
In this case one may begin with an initial spin network
$\Gamma_0$ with a finite number of edges and nodes
(This corresponds to the volume of space being finite.)  
One then has a finite procedure that constructs
a finite set of possible successor spin networks 
$\Gamma_1^\alpha$, where $\alpha$ labels the different
possibilities.  To each of these the theory associates a
quantum amplitude 
${\cal A}_{\Gamma_0 \rightarrow \Gamma_1^\alpha}$.

The procedure may then be applied to each of these,
producing a new set $\Gamma_2^{\alpha \beta}$.
Here $\Gamma_2^{\alpha \beta}$ labels the possible successors to
each of the $\Gamma_1^\alpha$.  The procedure may
be iterated any finite number of times $N$, producing a set
of spin networks ${\cal S}^N_{\Gamma_0}$
that grow out of the initial spin network
$\Gamma_0$ after $N$ steps.  ${\cal S}^N_{\Gamma_0}$ is
itself a directed graph, where two spin networks are joined
if one is a successor of the other.  There may be more than
one path in ${\cal S}^N_{\Gamma_0}$ between 
$\Gamma_0$ and some spin network $\Gamma_{final}$.
The amplitude for $\Gamma_0$ to evolve to
$\Gamma_{final}$ is then the sum over the paths
that join them in ${\cal S}^N_{\Gamma_0}$, in the
limit $N \rightarrow \infty$ of the products of the amplitudes
for each step along the way.

For any finite $N$, ${\cal S}^N_{\Gamma_0}$ has a finite number
of elements and the procedure is finitely specifiable.  There may
be issues about taking the limit $N \rightarrow \infty$, but
there is no reason to think that they are worse than similar
problems in quantum mechanics or 
quantum field theory.  
In any case, there is a sense in which each step takes a certain
amount of time,  in the limit $N \rightarrow \infty$ we will
be picking up the probability amplitude for the transition to
happen in infinite time.

Each step represents a finite time evolution because it corresponds
to certain causal processes by which information is
propagated in the spin network.  The rule by which the
amplitude is specified satisfies a principle of causality,
by which information about an element of a successor
network only depends on a small region of the its predecessor.
There are then discrete analogues of light cones and causal
structures in the theory.  Because the geometry associated
to the spin networks is discrete\cite{volume1}, the process by which
information at two nearby nodes or edges may propagate to
jointly influence the successor network is finite, not infinitesimal.

In ordinary quantum systems
it is usually the case that 
there is a non-vanishing probability for a state to evolve to an 
infinite number of elements of a basis after a finite amount of time.
The procedure we've just described 
then differs from ordinary quantum mechanics,
in that there are a finite number of possible successors for each
basis state after a finite evolution.    The reason is again causality
and discreteness: since the spin networks represent discrete
quantum geometries, and since information must only flow
to neighboring sites of the graph in a finite series of steps, at
each elementary step there are only a finite number of things
that can happen.

We may note that if the Hilbert space is not
constructible, we cannot ask if this procedure is unitary.
But we can still normalize the amplitudes so that the
sum of the absolute squares of the amplitudes to evolve
from any spin network to its successors is unity.  This gives
us something weaker than unitarity, but strong enough to
guarantee that probability  is conserved locally in the
space of configurations.

To summarize, in such an approach, the amplitude to evolve from the
initial spin network $\Gamma_0$ to any element
of   ${\cal S}^N_{\Gamma_0}$, for large finite $N$ is 
computable, even if it is the case that the spin networks
cannot be classified so that the basis itself
is not finitely specifiable.  Thus, such a procedure gives a
way to do quantum physics even for cases in which the
Hilbert space is not constructible.  

We may make two comments about this form of resolution
of the problem.  First, it necessarily involves an element of
time and causality.  The way in which the amplitudes are
constructed in the absence of a specifiable basis or
Hilbert structure requires a notion of successor states.  
The theory never has to ask about the whole space of
states, it only explores a finite set of successor states at each
step.  Thus, a notion of time is necessarily introduced.

Second, we might ask how we might formalize such a theory.
The role of the space of all states is replaced by the notion of
the successor states of a given network.  The immediate 
successors to a graph $\Gamma_0$ may be called the
{\it adjacent possible}\cite{investigations}. They are finite in number and
constructible.  They replace the idealization of all possible
states that is used in ordinary quantum mechanics.  We  may
note a similar notion of an {\it adjacent possible} set of
configurations, reachable from a given configuration in one
step, plays a role in formalizations of the self-organization
of biological and other complex systems\cite{investigations}.

In such a formulation there is no need to construct the state space
{\it a priori}, or equip it with a structure such as an inner product.
One has simply a set of rules by which a set of possible configurations
and histories of the universe is constructed by a finite procedure,
given any initial state.  In a sense it may be said that the system
is constructing the space of its possible states and histories as it
evolves.  

Of course, were we to do this for all initial states, we would have
constructed the entire state space of the theory.  But there are an
infinite number of possible initial states and, as we have been arguing,
they may not be classifiable.  In this case it is the evolution itself
that constructs the subspace of the space of states that is needed
to describe the possible futures of any given state.  And by doing so
the construction gives us an intrinsic notion of time.

\section{Conclusions}

We must emphasize first of all that these comments are meant to
be preliminary.  Their ultimate relevance rests partly on the
issue of whether there is a decision procedure for spin networks
(or perhaps for some extension of them that turns out to be relevant for
real quantum gravity\cite{future}.).  But more importantly, it 
suggests an alternative
type of framework for constructing quantum theories of cosmology, in
which there is no {\it a priori} configuration space or Hilbert 
space structure, but in which the theory is defined entirely in terms
of the sets of adjacent possible configurations, accessible from any
given configuration.  Whether such formulations turn out to be
successful at resolving all the problems of quantum gravity and
cosmology is a question that must 
be left for the future\footnote{We may note that 
the notion of an evolving Hilbert
space structure may be considered apart from the issues discussed
here\cite{rodolfo-evolving}.}.

There are further implications for theories of cosmology, if it turns out to
be the case that their configuration space or state space is not
finitely constructible.  One is to the problem of whether the second law
of thermodynamics applies at a cosmological scale. If the configuration
space or state space is not constructible, then it is not clear that the
ergodic hypothesis is well defined or useful.  Neither may the standard
formulations of statistical mechanics be applied.  What is then needed
is a new approach to statistical physics based only on the evolving
set of possibilities generated by the evolution from a given initial
state.  It is possible to speculate whether there may in such a context
be a ``fourth law" of thermodynamics in which the evolution 
extremizes the dimension of the adjacent possible, which is 
the set of states accessible to the
system at any stage in its evolution\cite{investigations}.

Finally, we may note that there are other reasons to suppose
that a quantum cosmological theory must incorporate some  
mechanisms analogous to the  self-organization of complex 
systems\cite{book}.
For example, these may be necessary to
tune the system to the critical behavior necessary for the existence
of the classical limit\cite{critical,causal}.  
This may also be necessary if the universe
is to have sufficient complexity that a four manifolds worth
of spacetime events are completely distinguished by purely 
relational observables\cite{spaceandtime,book}. The arguments given
here are complementary to those, and provide yet another way in which
notions of self-organization may play a role in a fundamental
cosmological theory.

\section*{ACKNOWLEDGEMENTS}

We are indebted to Julian Barbour and Fotini
Markopoulou for conversations which were very helpful in formulating
these ideas. We would also like to thank John Baez, Louis Crane, 
Lou Kauffman
and Adrian Ocneanu for discussions and 
help concerning the mathematical questions
about classifiability.   This work was 
supported by NSF grant PHY-9514240 to The Pennsylvania State
University, a NASA grant to The Santa Fe Institute. 
Finally, we are grateful to the
organizers of the conference on Fundamental Sources of
Unpredictability for providing the opportunity of
beginning discussions that led to this paper.

\end{document}